\documentclass[twocolumn,tighten]{aastex631}

\usepackage{natbib}
\usepackage{multirow}
\usepackage{amsmath}
\usepackage{hyperref}
\hypersetup{citecolor=blue,urlcolor=blue}

\shorttitle{ALMA observations of the $z = 6.05$ quasar-quasar pair system}
\shortauthors{T. Izumi et al.}
\begin{document}
\title{Merging gas-rich galaxies that harbor low-luminosity twin quasars at $z = 6.05$: \\
a promising progenitor of the most luminous quasars}

\correspondingauthor{Takuma Izumi}
\email{takuma.izumi@nao.ac.jp}

\author[0000-0001-9452-0813]{Takuma Izumi}
\affil{National Astronomical Observatory of Japan, 2-21-1 Osawa, Mitaka, Tokyo 181-8588, Japan}
\affil{Department of Astronomical Science, The Graduate University for Advanced Studies, SOKENDAI, 2-21-1 Osawa, Mitaka, Tokyo 181-8588, Japan}
\affil{Amanogawa Galaxy Astronomy Research Center, Kagoshima University, 1-21-35 Korimoto, Kagoshima 890-0065, Japan} 
\author[0000-0001-5063-0340]{Yoshiki Matsuoka} 
\affil{Research Center for Space and Cosmic Evolution, Ehime University, 2-5 Bunkyo-cho, Matsuyama, Ehime 790-8577, Japan} 
\author[0000-0003-2984-6803]{Masafusa Onoue} 
\altaffiliation{Kavli Astrophysics Fellow}
\affil{Kavli Institute for the Physics and Mathematics of the Universe (Kavli IPMU, WPI), The University of Tokyo Institutes for Advanced Study, The University of Tokyo, Kashiwa, Chiba 277-8583, Japan} 
\affil{Center for Data-Driven Discovery, Kavli IPMU (WPI), UTIAS, The University of Tokyo, Kashiwa, Chiba 277-8583, Japan}
\affil{Kavli Institute for Astronomy and Astrophysics, Peking University, Beijing 100871, P.R.China}
\author[0000-0002-0106-7755]{Michael A. Strauss}
\affil{Princeton University Observatory, Peyton Hall, Princeton, NJ 08544, USA}
\author[0000-0003-1937-0573]{Hideki Umehata}
\affil{Institute for Advanced Research, Nagoya University, Furocho, Chikusa, Nagoya 464-8602, Japan}
\affil{Department of Physics, Graduate School of Science, Nagoya University, Furocho, Chikusa, Nagoya 464-8602, Japan}
\affiliation{Cahill Center for Astronomy and Astrophysics, California Institute of Technology, 1200 E California Blvd, MC 249-17, Pasadena, CA
91125, USA}
\author[0000-0002-0000-6977]{John D. Silverman}
\affil{Kavli Institute for the Physics and Mathematics of the Universe (Kavli IPMU, WPI), The University of Tokyo Institutes for Advanced Study, The University of Tokyo, Kashiwa, Chiba 277-8583, Japan}
\affil{Center for Data-Driven Discovery, Kavli IPMU (WPI), UTIAS, The University of Tokyo, Kashiwa, Chiba 277-8583, Japan}
\affil{Department of Astronomy, School of Science, The University of Tokyo, 7-3-1 Hongo, Bunkyo, Tokyo 113-0033, Japan}
\affil{Center for Astrophysical Sciences, Department of Physics \& Astronomy, Johns Hopkins University, Baltimore, MD 21218, USA} 
\author[0000-0002-7402-5441]{Tohru Nagao}
\affil{Research Center for Space and Cosmic Evolution, Ehime University, 2-5 Bunkyo-cho, Matsuyama, Ehime 790-8577, Japan}
\affil{Amanogawa Galaxy Astronomy Research Center, Kagoshima University, 1-21-35 Korimoto, Kagoshima 890-0065, Japan} 
\author[0000-0001-6186-8792]{Masatoshi Imanishi}
\affil{National Astronomical Observatory of Japan, 2-21-1 Osawa, Mitaka, Tokyo 181-8588, Japan}
\affil{Department of Astronomical Science, The Graduate University for Advanced Studies, SOKENDAI, 2-21-1 Osawa, Mitaka, Tokyo 181-8588, Japan}
\author[0000-0002-4052-2394]{Kotaro Kohno}
\affil{Institute of Astronomy, Graduate School of Science, The University of Tokyo, 2-21-1 Osawa, Mitaka, Tokyo 181-0015, Japan}
\affil{Research Center for the Early Universe, Graduate School of Science, The University of Tokyo, 7-3-1 Hongo, Bunkyo-ku, Tokyo 113-0033, Japan}
\author[0000-0002-3531-7863]{Yoshiki Toba}
\affil{National Astronomical Observatory of Japan, 2-21-1 Osawa, Mitaka, Tokyo 181-8588, Japan}
\affil{Academia Sinica Institute of Astronomy and Astrophysics, 11F of Astronomy-Mathematics Building, AS/NTU, No.1, Section 4, Roosevelt Road, Taipei 10617, Taiwan}
\affil{Research Center for Space and Cosmic Evolution, Ehime University, 2-5 Bunkyo-cho, Matsuyama, Ehime 790-8577, Japan}
\author[0000-0002-4923-3281]{Kazushi Iwasawa}
\affil{Institut de Ci\`encies del Cosmos (ICCUB), Universitat de Barcelona (IEEC-UB), Mart\'i i Franqu\`es, 1, 08028 Barcelona, Spain}
\affil{ICREA, Pg. Llu\'is Companys 23, 08010 Barcelona, Spain}
\author[0000-0002-6939-0372]{Kouichiro Nakanishi}
\affil{National Astronomical Observatory of Japan, 2-21-1 Osawa, Mitaka, Tokyo 181-8588, Japan}
\affil{Department of Astronomical Science, The Graduate University for Advanced Studies, SOKENDAI, 2-21-1 Osawa, Mitaka, Tokyo 181-8588, Japan}
\author[0009-0003-5438-8303]{Mahoshi Sawamura}
\affil{Department of Astronomy, School of Science, The University of Tokyo, 7-3-1 Hongo, Bunkyo, Tokyo 113-0033, Japan}
\affil{National Astronomical Observatory of Japan, 2-21-1 Osawa, Mitaka, Tokyo 181-8588, Japan}
\author[0000-0001-7201-5066]{Seiji Fujimoto}
\affil{Department of Astronomy, The University of Texas at Austin, Austin, TX 78712, USA}
\affil{Cosmic Dawn Center (DAWN), Copenhagen, Denmark}
\affil{Niels Bohr Institute, University of Copenhagen, Jagtvej 128, 2200 Copenhagen N}
\author[0000-0003-3214-9128]{Satoshi Kikuta}
\affil{Department of Astronomy, School of Science, The University of Tokyo, 7-3-1 Hongo, Bunkyo, Tokyo 113-0033, Japan}
\author[0000-0002-3866-9645]{Toshihiro Kawaguchi}
\affil{Department of Economics, Management and Information Science, Onomichi City University, Hisayamada 1600-2, Onomichi, Hiroshima 722-8506, Japan} 
\author[0000-0003-4569-1098]{Kentaro Aoki}
\affil{Subaru Telescope,  National Astronomical Observatory of Japan}
\author[0000-0002-6821-8669]{Tomotsugu Goto} 
\affil{Institute of Astronomy, National Tsing Hua University, No.101, Section 2, Kuang-Fu Road, Hsinchu 30013, Taiwan0000-0002-6821-8669}

\begin{abstract}
We present ALMA [\ion{C}{2}] 158 $\micron$ line and underlying far-infrared (FIR) continuum emission observations 
($0''.57 \times 0''.46$ resolution) toward a quasar-quasar pair system recently discovered at $z = 6.05$ (Matsuoka et al. 2024). 
The quasar nuclei (C1 and C2) are faint ($M_{\rm 1450} \gtrsim -23$ mag), 
but we detect very bright [\ion{C}{2}] emission bridging the 12 kpc between the two objects and extending beyond them (total luminosity $L_{\rm [CII]} \simeq 6 \times 10^9~L_\odot$). 
The [\ion{C}{2}]-based total star formation rate of the system is $\sim 550~M_\odot$ yr$^{-1}$ 
(IR-based dust-obscured SFR is $\sim 100~M_\odot$ yr$^{-1}$), 
with a [\ion{C}{2}]-based total gas mass of $\sim 10^{11}~M_\odot$. 
The dynamical masses of the two galaxies are large ($\sim 9 \times 10^{10}~M_\odot$ for C1 and $\sim 5 \times 10^{10}~M_\odot$ for C2). 
There is a smooth velocity gradient in [\ion{C}{2}], indicating that these quasars are a tidally interacting system. 
We identified a dynamically distinct, fast [\ion{C}{2}] component around C1: 
detailed inspection of the line spectrum there reveals the presence of a broad wing component, 
which we interpret as the indication of fast outflows with a velocity of $\sim 600$ km s$^{-1}$. 
The expected mass loading factor of the outflows, after accounting for multiphase gas, is $\gtrsim 2-3$, which is intermediate between AGN-driven and starburst-driven outflows. 
Hydrodynamic simulations in the literature predicted that this pair will evolve to a luminous ($M_{\rm 1450} \lesssim -26$ mag), 
starbursting ($\gtrsim 1000~M_\odot$ yr$^{-1}$) quasar after coalescence, one of the most extreme populations in the early universe. 
\end{abstract}

\section{Introduction}\label{sec1} 
In the hierarchical structure formation scenario, galaxies undergo multiple mergers over cosmic time. 
Models show that mergers of gas-rich galaxies trigger intense star formation 
and fueling onto the central supermassive black holes (SMBHs), which appear as luminous quasars 
\citep[e.g.,][]{1988ApJ...325...74S,2005Natur.433..604D,2006ApJS..163....1H}. 
Some theoretical models predict that subsequent feedback from the quasars, or active galactic nuclei (AGNs), plays 
a crucial role in driving the co-evolution of SMBHs and host galaxies \citep{2015ARA&A..53..115K,2020A&ARv..28....2V}, leading to the observed tight correlation 
between the masses of SMBHs ($M_{\rm BH}$) and those of the host galaxy bulges observed in the local universe \citep{2013ARA&A..51..511K}. 
Detections of galaxy-scale AGN-driven outflows in multiphase gas 
\citep[e.g.,][]{2008A&A...491..407N,2012A&A...537A..44A,2014A&A...562A..21C}, 
a higher AGN fraction in interacting/merging systems \citep[e.g., ][]{2011ApJ...743....2S,2018PASJ...70S..37G,2018Natur.563..214K}, 
and the global similarity in star-formation and SMBH accretion histories over cosmic time 
\citep{2014ARA&A..52..415M}, support this scenario. 

It is intriguing in this context that massive ($\sim 10^{11}~M_\odot$), quiescent, and old galaxies 
are already formed at $z \sim 4-5$ \citep[e.g.,][]{2023Natur.619..716C,2023ApJ...947...20V}, 
suggesting that a phase of rapid growth of galaxies and SMBHs, and their associated feedback, had happened at even higher redshifts. 
Indeed, more than 400 quasars with rest-UV magnitude of $M_{\rm 1450} < -22$ mag are known at $z > 5.7$ to date \citep{2020ARA&A..58...27I,2023ARA&A..61..373F}, 
most of which have been identified by wide-field optical and near-infrared surveys 
\citep[e.g.,][]{2016ApJ...833..222J,2016ApJS..227...11B,2016ApJ...828...26M,2018PASJ...70S..35M,2018ApJS..237....5M}. 
Sub/millimeter observations of the rest-frame far-infrared (FIR) continuum and C$^+$ $^2P_{3/2}$ $\rightarrow$ $^2P_{1/2}$ 157.74 $\micron$ line 
([\ion{C}{2}] 158 $\micron$) emission, the latter is one of the prime coolants of the cold interstellar medium/ISM \citep{2022ARA&A..60..247W}, 
toward optically luminous quasars ($M_{\rm 1450} \lesssim -26$ mag) 
by the Atacama Large Millimeter/submillimeter Array (ALMA), have revealed 
that vigorous starburst (star formation rate SFR $\gtrsim 100-1000~M_\odot$ yr$^{-1}$) 
and huge amounts of dust ($\sim 10^8~M_\odot$) and gas ($\sim 10^{10}~M_\odot$) are usually associated with these quasars 
\citep[e.g.,][]{2013ApJ...773...44W,2016ApJ...816...37V,2020ApJ...904..130V,2022A&A...662A..60D}. 
Although the prevalence of massive AGN-driven outflows remains unclear from observations of the neutral ISM \citep{2019A&A...630A..59B,2020ApJ...904..131N,2024ApJ...962....1S}, 
fast ionized outflows (probed by, e.g., [\ion{O}{3}] 5007{\AA}) are frequently seen in $z > 6$ quasars with recent James Webb Space Telescope (JWST) observations 
\citep[e.g.,][]{2023A&A...678A.191M,2023arXiv230904614Y,2023ApJ...951L...5Y,2024arXiv240213319L}. 

While multiwavelength observations of quasars were progressed significantly in recent years, 
understanding of their progenitors lags behind. Some limited studies on partially dust-obscured quasars \citep{2022Natur.604..261F} 
and starburst galaxies \citep{2013Natur.496..329R,2018Natur.553...51M,2018NatAs...2...56Z}, 
both of which may represent earlier evolutionary phases than UV-bright quasar phase, 
indeed revealed very rapid mass assembly in these systems at $z > 6-7$. 
Closely interacting galaxies are considered to be an earlier evolutionary stage than these, 
yet there are very few examples known to host SMBHs at $z \gtrsim 6$ \citep{2021ApJ...921L..27Y,2023AJ....165..191Y}. 
For example, \citet{2019ApJ...882...10N} studied five pairs of quasar host-companion galaxy at [\ion{C}{2}], 
and identified evidence of tidal interaction in three of them. 
Some other works have also found companion galaxies around quasars at $z > 5-6$ \citep[e.g.,][]{2017Natur.545..457D,2020ApJ...904..130V}. 
\citet{2019ApJ...880..157D} performed high resolution [\ion{C}{2}] and FIR continuum observations 
of another quasar-galaxy pair PJ308$-$21 at $z = 6.23$ that hosts an SMBH of $\sim 3 \times 10^9~M_\odot$ \citep{2024arXiv240213319L}, 
revealing a large amount of cold ISM and the highly interacting nature of the system (two companions with a projected distance of $\sim 5$ and $\gtrsim 10$ kpc, respectively). 

However, the interacting quasars targeted in the above papers are intrinsically as luminous as those of the other, 
isolated luminous quasars at $z > 6$ (such as those discovered by SDSS), suggesting that the phase of active galaxy interaction had already happened. 
On the other hand, hydrodynamic simulations of mergers of galaxies predict that both SFR and quasar luminosity 
increase by orders of magnitude when gas-rich galaxies merge \citep[e.g.,][]{2006ApJS..163....1H}. 
Thus, progenitors in the earlier interaction phase, that will evolve to the luminous quasars currently observed, are anticipated to be much fainter. 
Wide-field optical deep imaging surveys such as the Subaru Hyper Suprime-Cam (HSC) Strategic Survey \citep{2018PASJ...70S...8A} are useful to search for such faint objects. 
Indeed, we have established a multi-wavelength follow-up consortium for $z \gtrsim 6$ quasars discovered by the HSC survey, 
the Subaru High-z Exploration of Low-Luminosity Quasars \citep[SHELLQs, e.g.,][]{2016ApJ...828...26M,2018PASJ...70S..35M,2018ApJS..237....5M,2019ApJ...883..183M,2022ApJS..259...18M}, 
and so far discovered $> 150$ low-luminosity quasars down to $M_{\rm 1450} \sim -22$ mag at $z > 6$.

\subsection{Our target: a pair of quasars at z = 6.05}\label{sec1.1}
This paper presents our ALMA Cycle 7 observations of the [\ion{C}{2}] 158 $\micron$ line 
and the underlying rest-frame FIR continuum emission toward the twin quasar system at $z = 6.05$ discovered by our Subaru HSC survey, 
namely HSC $J$121503.42$-$014858.7 and HSC $J$121503.55$-$014859.3 (C1 and C2, hereafter), separated by 12 kpc projected on the sky \citep{Matsuoka2024}. 
This is the most distant quasar pairs (projected separation $\lesssim 10$ kpc, $\Delta z \lesssim 0.01$) known to date: 
the discovery and rest-UV properties of these quasars are reported in our companion paper \citep{Matsuoka2024}. 
The presence of pair quasars is a natural consequence of the hierarchical structure formation, 
yet the sample is very limited due to the rarity of SMBHs and the relatively short duration of the AGN phase. 

The two quasars in this study have similar rest-UV properties with $M_{\rm 1450} = -23.11$ mag (C1) and $M_{\rm 1450} = -22.66$ mag (C2): 
these correspond to quasar bolometric luminosities ($L_{\rm Bol}$) of $6.2 \times 10^{45}$ erg s$^{-1}$ (C1) and $4.1 \times 10^{45}$ erg s$^{-1}$ (C2). 
The UV magnitudes of these are roughly 10 times higher than the knee of the galaxy luminosity function at $z \sim 6$ \citep{2022ApJS..259...20H}, 
but are fainter than the characteristic value of the $z \sim 6$ quasar luminosity function \citep[$M_{\rm 1450} = -24.9$ mag,][]{2018ApJ...869..150M}. 
The quasar nature of C1 + C2 has been further confirmed by the detections of broad Ly$\alpha$ (FWHM $> 1000$ km s$^{-1}$) in both objects with high luminosities ($>10^{43}$ erg s$^{-1}$). 
The broad Ly$\alpha$ components are sensitive to $M_{\rm BH}$ \citep{2024ApJ...960..112T}; 
\citet{Matsuoka2024} estimated their masses to be $\log (M_{\rm BH}/M_\odot) \simeq 8.1$ each, corresponding to Eddington ratios of $\sim 0.4$ (C1) and $\sim 0.3$ (C2). 

Throughout this work, we adopt the concordant Lambda Cold Dark Matter ($\Lambda$CDM) 
cosmology with $H_0$ = 70 km s$^{-1}$ Mpc$^{-1}$, 
$\Omega_{\rm M} = 0.3$, and $\Omega_{\rm \Lambda} = 0.7$. 
At the redshift of the source ($z = 6.05$), the age of the universe is 0.905 Gyr 
and an angular size of $1\arcsec$ corresponds to a proper distance of 5.68 kpc.

\section{ALMA Observations}\label{sec2} 
Our observations were conducted as an ALMA Director's Discretionary (DDT) program (2019.A.00019.S, PI = T. Izumi). 
There were two executions in ALMA Band 6 (centered on $\lambda_{\rm obs} = 1.1$ mm) on 2020 Feb 29, 
in a $\sim 23\arcsec$ diameter field of view with 41 antennas. 
Three spectral windows (each $\sim 1.875$ GHz wide) were placed 
on one side-band to maximize the contiguous frequency coverage. 
We set the phase-tracking center of this pointing to 
($\alpha_{\rm ICRS}$, $\delta_{\rm ICRS}$) = (12$^{\rm h}$15$^{\rm m}$03$^{\rm s}$.420, $-$01\arcdeg48\arcmin58\arcsec.70), 
which corresponds to the optical position of C1. 
The baseline length ranged from 15.1 m to 783.5 m, 
resulting in a maximum recoverable scale of $\sim 5\arcsec$. 
Two quasars, J1256$-$0547 and J1217$-$0029, were observed as a flux/bandpass calibrator and a phase calibrator, respectively. 
The total on-source time was 106 minutes. 

The data were first pipeline-processed using \verb|CASA| \citep{2007ASPC..376..127M,2022PASP..134k4501C} v.5.6 and further analyzed with v.6.1. 
All images presented in this work were reconstructed using the \verb|tclean| task down to $3\sigma$ level, 
with Briggs weighting (robust = 2.0) to maximize the point source sensitivity. 
For the [\ion{C}{2}] cube, we averaged several channels to obtain a velocity resolution of 50 km s$^{-1}$, 
which resulted in a 1$\sigma$ channel sensitivity of 0.14 mJy beam$^{-1}$ (beam size = $0\arcsec.57 \times 0\arcsec.46$, P.A. = $-41\arcdeg.2$). 
Line-free channels were integrated to generate a continuum map 
($0\arcsec.56 \times 0\arcsec.45$, P.A. = $-43\arcdeg.1$, 1$\sigma$ = 17.2 $\mu$Jy beam$^{-1}$), 
which we subtracted in the $uv$ plane using the task \verb|uvcontsub| with a first-order polynomial function, before making the line cube. 
This continuum map was made with the multi-frequency synthesis (MFS) mode of the \verb|tclean| task. 
In this paper, we show only statistical errors unless mentioned otherwise. 
The absolute flux uncertainty is $\sim 10\%$ (ALMA Cycle 7 Proposer's Guide). 
We also used the \verb|MIRIAD| software \citep{1995ASPC...77..433S} for some analyses.

\section{Results}\label{sec3}
\subsection{A galaxy merger caught in the act}\label{sec3.1}
\begin{figure*}
\begin{center}
\includegraphics[width=\linewidth]{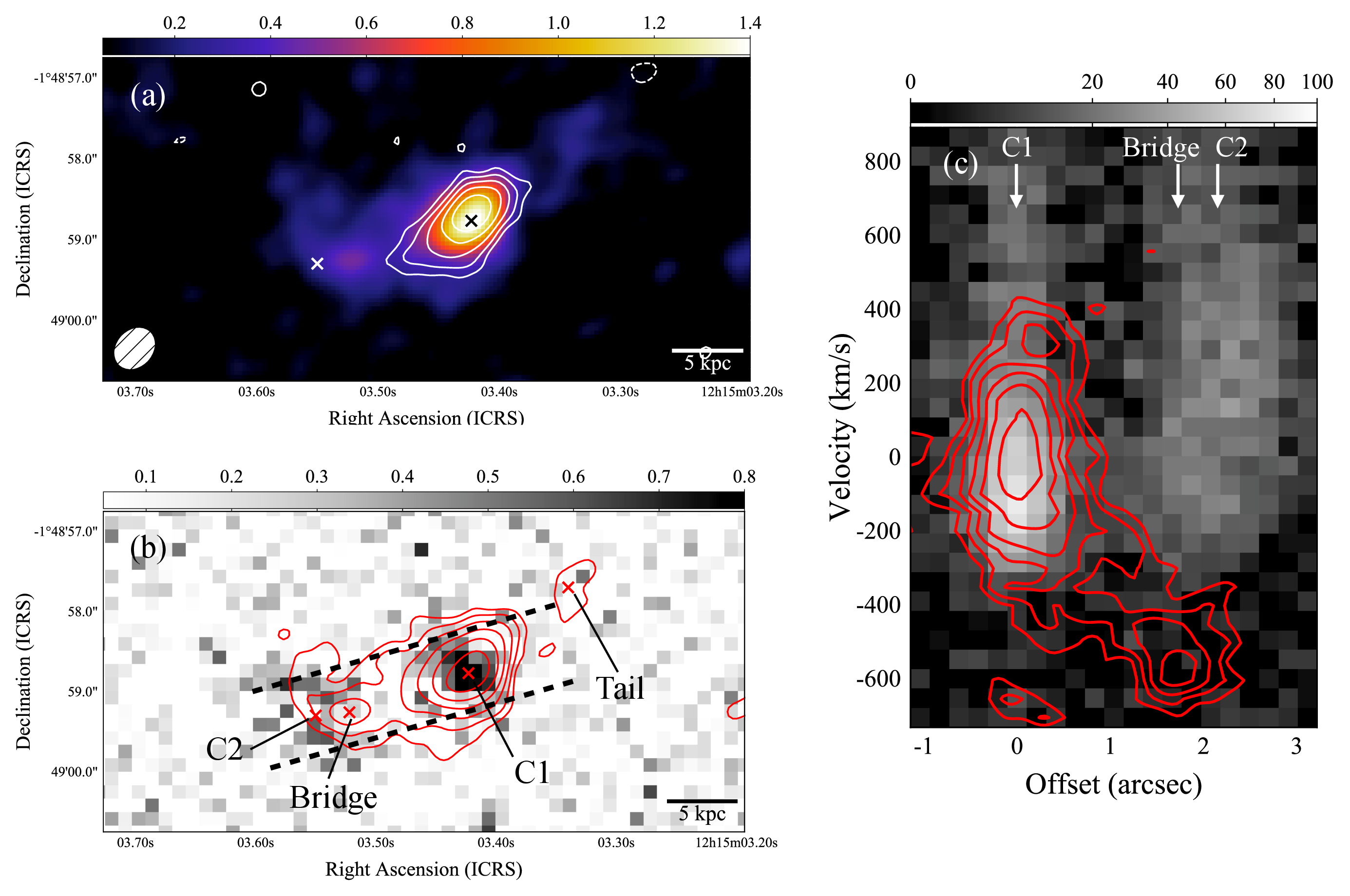}
\caption{
(a) Spatial distributions of the velocity-integrated (i.e., moment 0) [\ion{C}{2}] (color; Jy beam$^{-1}$ km s$^{-1}$ unit, $1\sigma = 0.039$ Jy beam$^{-1}$ km s$^{-1}$) 
and the rest-FIR continuum emission (contours) of the pair system. 
Contours start at $\pm 3\sigma$ ($1\sigma = 17.2$ $\mu$Jy beam$^{-1}$) and increase by factors of $\sqrt{2}$. 
Crosses mark the locations of C1 and C2 (the latter is determined as the flux-weighted center of the HSC $z$-band map). 
(b) Overlay of the [\ion{C}{2}] moment 0 map (contours) on the HSC $z$-band map (gray scale in arbitrary units). 
Contours start at $\pm 5\sigma$ and increase by factors of $\sqrt{2}$.
Notable features including C1 and C2 are marked by the crosses. 
The pseudo-slit used to extract the position-velocity diagram (PVD) in the panel-(c) is also marked 
(length $\sim 4''$, width $= 1''$, PA = 106$\arcdeg$ measured counter-clockwise from north). 
(c) PVD of Ly$\alpha$ \citep{Matsuoka2024} in gray scale (arbitrary unit) 
and that of [\ion{C}{2}] in contours (start from $2\sigma$ and increase by factors of $\sqrt{2}$, $1\sigma = 0.14$ mJy beam$^{-1}$). 
The position of C1 and its [\ion{C}{2}] line center velocity are set to the origin of the coordinates. 
The spatial locations of C1, C2, Bridge, are marked. 
}
\label{fig1}
\end{center}
\end{figure*}

We detected the [\ion{C}{2}] emission from both C1 and C2, and FIR continuum emission mainly from C1, as shown in Figure \ref{fig1}a. 
The emission properties as well as relevant physical quantities are summarized in Table \ref{tbl1}. 
It is noteworthy that the location of C1 (defined from the HSC $z$-band map; Figure \ref{fig1}b) precisely coincides 
with the peak position of the [\ion{C}{2}] velocity-integrated intensity map, confirming the astrometric accuracy of our observations. 
However, we found that the location of C2 is offset from the nearest local maximum of the [\ion{C}{2}] emission by $0''.42$ or 2.4 kpc. 
This local maximum lies between C1 and C2, and we refer to it as the $``$Bridge$"$ hereafter. 
Note that the location of C2 is determined as the flux-weighted center of the HSC $z$-band map. 
As the $z$-band emission is spatially extended, this location does not necessarily correspond to the exact location of the quasar. 
However, it is evident that the bright $z$-band emitting region is offset from this Bridge. 
We also found significant [\ion{C}{2}] emission north-west of C1 (offset by $\sim 1''.6$ or 9.3 kpc), which we refer to the $``$Tail$"$ in this work. 
Both the Bridge and the Tail are most likely formed by interactions of the host galaxies of C1 and C2. 

Figure \ref{fig1}c compares the position-velocity diagram (PVD) of Ly$\alpha$ \citep{Matsuoka2024} and that of [\ion{C}{2}]. 
It is also clear from this comparison that C2 (bright in Ly$\alpha$) and the Bridge (bright in [\ion{C}{2}]) are spatially offset. 
In addition, it is evident that the [\ion{C}{2}] redshift\footnote{The [C II] redshifts of C1 and C2 have been slightly updated from the values we reported in \citet{Matsuoka2024}.} 
of C2 is blueshifted relative to the Ly$\alpha$ redshift reported in \citet{Matsuoka2024} by $z_{\rm [CII]} - z_{\rm Ly\alpha} = -0.008$ (or $\sim -400$ km s$^{-1}$). 
This velocity offset would be due at least partly to strong Ly$\alpha$ absorption by the intergalactic medium. 
Indeed, this velocity offset lies nearly at the highest values found in comparably high redshift galaxies, 
and those galaxies with high velocity offset tend to show low Ly$\alpha$ escape fractions \citep{2020A&A...643A...6C}. 
Another plausible explanation would be that the Bridge is actually the satellite galaxy itself, 
and its quasar nucleus (C2) is located at the outskirt of this galaxy. 
In this case, C2 may have been kicked-out from the host galaxy (Bridge) during this interaction event with the main galaxy (host of C1), 
which have caused the velocity offset between Ly$\alpha$ and [\ion{C}{2}] as well. 
Future JWST observations of the stellar light distribution will allow us to assess whether this offset is 
due to a real displacement of the quasar from the gravitational center of the host galaxy. 

\begin{figure}
\begin{center}
\includegraphics[width=\linewidth]{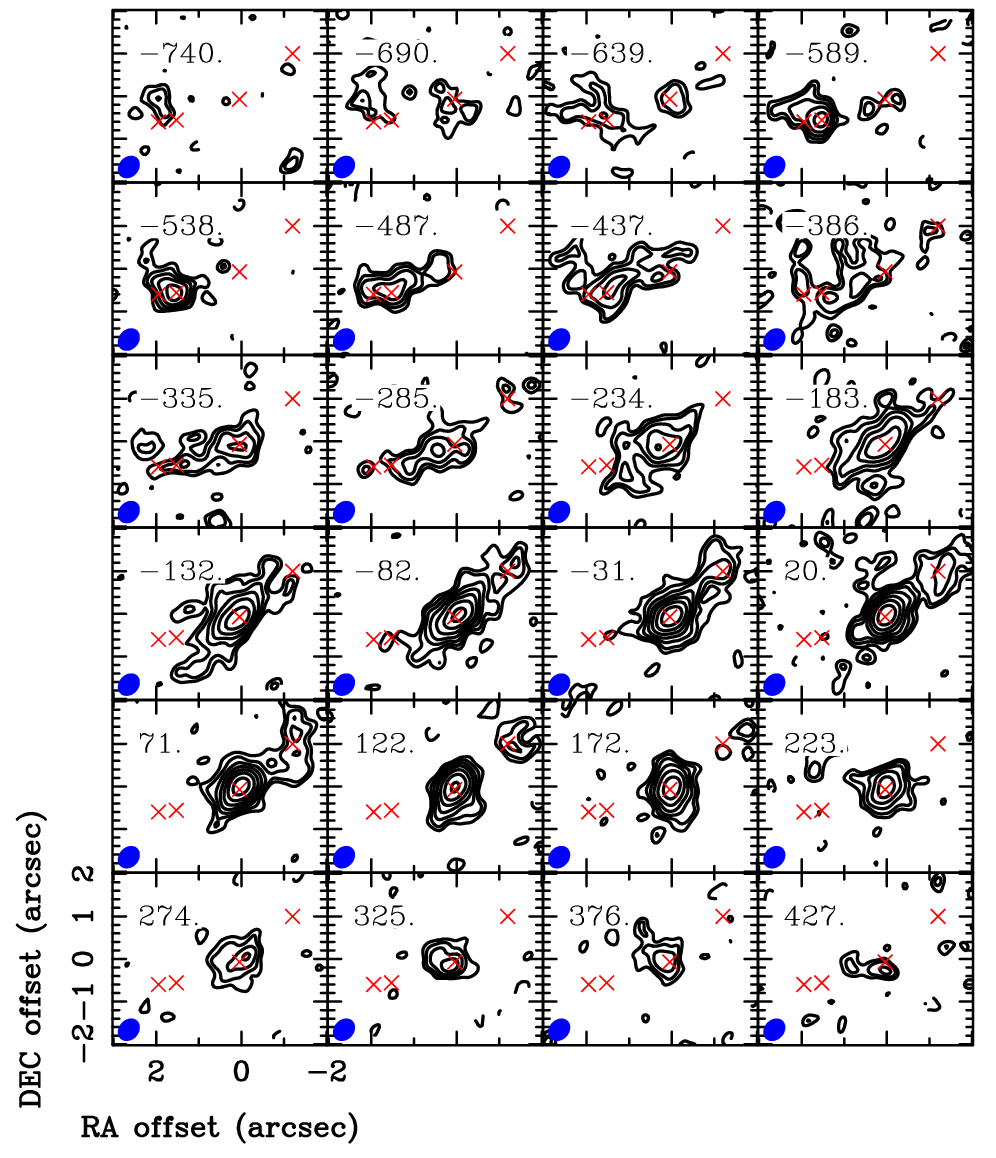}
\caption{
Velocity channel maps of the [\ion{C}{2}] line emission. 
Each channel is labeled with its central velocity in km s$^{-1}$ relative to the Gaussian centroid of the C1 line profile. 
The locations of C1 (the origin of the relative coordinates), C2, Bridge, and Tail are marked by the crosses in each panel (see also Figure \ref{fig1}). 
Contours start at $2\sigma$ with an increment of $\sqrt{2}$, where 1$\sigma$ = 0.14 mJy beam$^{-1}$. 
The synthesized beam is shown in the bottom left corner of each panel. 
}
\label{fig2}
\end{center}
\end{figure}

Figure \ref{fig2} shows the velocity channel maps of [\ion{C}{2}] emission. 
There is a clear velocity gradient from C2 to the Tail through the Bridge and C1, on scales of $\sim 4''$ or 23 kpc. 
This clearly indicates that the pair of galaxies are dynamically coupled. 
Thus, these galaxies will merge in a short time \citep[$< 1$ Gyr, e.g.,][]{2008ApJS..175..356H} to form a single massive galaxy. 
In addition to this, we found bright [\ion{C}{2}] emission at C1 in multiple channels of $>170$ km s$^{-1}$, 
suggesting a presence of dynamically distinct fast component there (see \S~\ref{sec3.4}).

\begin{deluxetable*}{c|ccc}
\tabletypesize{\small}
\tablecaption{Rest-FIR Properties of the Pair System \label{tbl1}}
\tablewidth{0pt}
\tablehead{
 & C1 & C2 + Bridge & Tail 
}
\startdata 
R.A. (ICRS) & 12$^{\rm h}$15$^{\rm m}$03$^{\rm s}$.423 & 12$^{\rm h}$15$^{\rm m}$03$^{\rm s}$.536 & 12$^{\rm h}$15$^{\rm m}$03$^{\rm s}$.340 \\ 
Dec (ICRS) & $-$01\arcdeg48\arcmin58\arcsec.77 & $-$01\arcdeg48\arcmin59\arcsec.28 & $-$01\arcdeg48\arcmin57\arcsec.70 \\ \hline
 & \multicolumn{3}{c}{[\ion{C}{2}] Emission} \\ \hline
$z_{\rm [CII]}$ & 6.0561 $\pm$ 0.0002 & 6.0447 $\pm$ 0.0003 &  6.0566 $\pm$ 0.0003 \\ 
Velocity offset (km s$^{-1}$) & 0 & $-481 \pm 13$ & $22 \pm 14$ \\ 
FWHM (km s$^{-1}$) & $594 \pm 20$ & $361 \pm 30$ & $317 \pm 34$ \\ 
$S_{\rm [CII]}\Delta V$ (Jy km s$^{-1}$) & $3.67 \pm 0.19$ & $0.91 \pm 0.13$ & $0.52 \pm 0.13$ \\
$L_{\rm [CII]}$ (10$^9$ $L_\odot$) & $3.50 \pm 0.18$ & $0.87 \pm 0.12$ & $0.49 \pm 0.12$ \\ 
SFR$_{\rm [CII]}$ ($M_\odot$ yr$^{-1}$) & $305 \pm 16$ & $75 \pm 11$ & $43 \pm 11$ \\ 
$M_{\rm H2}$ ($10^{10}~M_\odot$) & $\sim 10$ & $\sim 2.6$ & $\sim 1.5$ \\ \hline 
 & \multicolumn{3}{c}{Continuum Emission ($T_{\rm dust}$ = 35 K, $\beta$ = 1.6, $\kappa_\lambda = 0.77 ({\rm 850 \micron}/\lambda)^\beta$ cm$^2$ g$^{-1}$)} \\ \hline
$f_{\rm 1.1mm}$ (mJy) & $0.63 \pm 0.05$ & $< 0.09$ & $< 0.09$ \\
$L_{\rm FIR}$ (10$^{11}$ $L_\odot$) & $4.57 \pm 0.36$ & $< 0.71$ & $< 0.71$  \\
$L_{\rm TIR}$ (10$^{11}$ $L_\odot$) & $5.99 \pm 0.48$ & $< 0.94$ & $< 0.94$ \\
$M_{\rm dust}$ (10$^8$ $M_\odot$) & $1.58 \pm 0.13$ & $< 0.25$ & $< 0.25$ \\
SFR$_{\rm TIR}$ ($M_\odot$ yr$^{-1}$) & $89 \pm 7$ & $< 12$ & $< 12$ \\ 
\enddata
\tablecomments{Coordinates indicate the peak positions of the [\ion{C}{2}] emission for C1 and Tail. 
For C2, we adopt the flux-weighted centroid of the HSC $z$-band map \citep{Matsuoka2024}. 
This is spatially averaged with the position of the Bridge to define the position of C2 + Bridge. 
We used 1$''$.5 and 1$''$.0 diameter apertures to extract the emission properties of C1 and the other components, respectively. 
Upper limits indicate $3\sigma$ values. 
The redshifts were determined from single Gaussian fittings to the [\ion{C}{2}] spectra (see Appendix-\ref{App-I}), 
but the line fluxes were based on the integrated intensity map (Figure \ref{fig1}).}
\end{deluxetable*}

\subsection{Star-forming nature}\label{sec3.2}
With the [\ion{C}{2}] line and FIR continuum luminosities, we estimate basic star formation properties of this pair system (Table \ref{tbl1}). 
In order to encompass the emitting region sufficiently, we used a circular aperture of $1''.5$ diameter and $1''.0$ diameter respectively, 
to measure the fluxes of C1 and the remaining notable regions of C2 + Bridge, and Tail. 
Note that, as the spatial separation of C2 (12$^{\rm h}$15$^{\rm m}$03$^{\rm s}$.55, $-$01\arcdeg48\arcmin59\arcsec.3) 
and the Bridge (12$^{\rm h}$15$^{\rm m}$03$^{\rm s}$.522, $-$01\arcdeg48\arcmin59\arcsec.26) is only $0''.42$, we present the value at ``C2 + Bridge" by placing the aperture 
at the middle point (12$^{\rm h}$15$^{\rm m}$03$^{\rm s}$.536, $-$01\arcdeg48\arcmin59\arcsec.28) of C2 and the Bridge. 
The individual [\ion{C}{2}] spectra of these regions, as well as the velocity channel maps made 
with a smoothed $1''.0$ resolution, are presented in Appendix-\ref{App-I} and \ref{App-II}.

\subsubsection{From [\ion{C}{2}] emission}
Following \citet{2005ARA&A..43..677S}, the [\ion{C}{2}] line luminosity is calculated as 
\begin{equation}
L_{\rm [CII]}/L_\odot = 1.04 \times 10^{-3}~S_{\rm [CII]}\Delta V~\nu_{\rm rest}~(1+z)^{-1}~D^2_L,
\end{equation}
where $S_{\rm [CII]}\Delta V$ is the [\ion{C}{2}] line flux in Jy km s$^{-1}$, $\nu_{\rm rest}$ is the rest frequency of 1900.5369 GHz, 
and $D_L$ is the luminosity distance to the target. 
The line luminosities of these regions lie in the range $(0.5-3.5) \times 10^9~L_\odot$. 
The brightest [\ion{C}{2}] emission originates from C1 ($3.5 \times 10^9~L_\odot$), 
which is the highest [\ion{C}{2}] luminosity reported for HSC low-luminosity quasars to date that typically have $L_{\rm [CII]} < 10^9~L_\odot$ 
\citep{2018PASJ...70...36I,2019PASJ...71..111I,2021ApJ...908..235I,2021ApJ...914...36I}, 
and is as high as those of optically luminous quasars \citep[e.g.,][]{2016ApJ...816...37V,2020ApJ...904..130V,2018ApJ...854...97D}. 
This luminosity is also comparable to the luminosities of PJ308$-$21 \citep[a galaxy merger at $z = 6.23$,][]{2019ApJ...880..157D}, 
although C1 ($M_{\rm 1450} = -23.1$ mag) is $\sim 20\times$ fainter than the quasar nucleus of PJ308$-$21 ($M_{\rm 1450} = -26.3$ mag) at rest-UV wavelength. 
We speculate that we are witnessing the recent onset of AGN, which will evolve to a more luminous quasar, in our pair system. 

By further assuming that the [\ion{C}{2}] line is excited primarily by star formation, 
we can estimate the SFR using the \citet{2014A&A...568A..62D} calibration based on local \ion{H}{2}/starburst galaxies: 
$\log ({\rm SFR_{[CII]}}/M_\odot~{\rm yr^{-1}}) = -7.06 + 1.0 \times \log (L_{\rm [CII]}/L_\odot)$, with a factor of two calibration uncertainty. 
This relation is applicable to high redshift ($z \sim 4-8$) star-forming galaxies 
as demonstrated by \citet{2020A&A...643A...3S} and \citet{2020A&A...643A...1L}. 
With this, we find SFR$_{\rm [CII]} = 305 \pm 16~M_\odot$ yr$^{-1}$ in C1 and 
$75 \pm 11~M_\odot$ yr$^{-1}$ in C2 + Bridge. 
If some of the [\ion{C}{2}] excitation is due to the quasars, our derived SFRs are upper limits. 

The [\ion{C}{2}] line luminosity can be used to estimate the molecular gas mass ($M_{\rm H2}$) as well \citep{2020A&A...643A.141M}. 
Using the calibration of \citet{2018MNRAS.481.1976Z} for a collection of main sequence and starburst galaxies at intermediate redshifts 
($M_{\rm H2}/M_\odot \sim 30 \times L_{\rm [CII]}/L_\odot$), 
we found $M_{\rm H2} \sim 10, 2.6, 1.5 \times 10^{10}~M_\odot$ in the regions of C1, C2 + Bridge, and Tail, respectively. 
Note that $L_{\rm [CII]}$ has a better correlation with $M_{\rm H2}$ than SFR \citep{2022ApJ...929...92V}, which support a reliability of this line luminosity 
as a molecular gas mass tracer, and also suggest that the $L_{\rm [CII]}$--SFR relation may arise from the combination of $L_{\rm [CII]}$--$M_{\rm H2}$ relation 
and $M_{\rm H2}$--SFR relation \citep[e.g.,][]{2012ARA&A..50..531K}. 
Hence, the [\ion{C}{2}]-based $M_{\rm H2}$ and SFR$_{\rm [CII]}$ reported here are not necessarily independent.

The total [\ion{C}{2}] line flux measured with a $6''.0$ diameter aperture is 6.6 Jy km s$^{-1}$, 
which translates to $L_{\rm [CII]}$, SFR$_{\rm [CII]}$, and $M_{\rm H2}$ 
of $6.3 \times 10^9~L_\odot$, 550 $M_\odot$ yr$^{-1}$, and $1.9 \times 10^{11}~M_\odot$, respectively. 
These values are among the highest ones found in $z > 6$ quasars of all UV luminosities 
\citep[e.g.,][]{2010ApJ...714..699W,2017ApJ...845..154V,2022A&A...662A..60D}, 
demonstrating the presence of active star formation and an immense gaseous reservoir. 
Note that \citet{2022A&A...662A..60D} pointed out discrepancies between [\ion{C}{2}]-based $M_{\rm H2}$ 
and CO-based, [\ion{C}{1}]-based, and dust continuum-based $M_{\rm H2}$: [\ion{C}{2}]-based values are systematically greater than the others by a factor of $\sim 3-5$. 
Even so, we claim that the total molecular gas mass of this pair is on the order of $10^{10}~M_\odot$, 
which is enough to sustain high SFR (e.g., 500 $M_\odot$ yr$^{-1}$) and luminous quasar activity 
(mass accretion rate $\sim 10~M_\odot$ yr$^{-1}$) expected to happen after a galaxy merger over $\sim 10^8$ yrs.

\subsubsection{From FIR continuum emission}
The rest-FIR continuum emission mainly originates from the regions around C1 (Figure \ref{fig1}a). 
Our 2D Gaussian fit to the FIR continuum distribution using the CASA task \verb|imfit| found a peak position of 
(RA, Dec) = (12$^{\rm h}$15$^{\rm m}$03$^{\rm s}$.426, $-$01\arcdeg48\arcmin58\arcsec.73), 
which agrees well with the optical coordinates of C1 \citep{Matsuoka2024}. 
The beam-deconvolved size of this 2D Gaussian is $(1''.19 \pm 0''.14) \times (0''.54 \pm 0''.07)$ 
or (6.76 kpc $\pm$ 0.80 kpc) $\times$ (3.04 kpc $\pm$ 0.40 kpc), and the peak and area-integrated total 
flux densities are 220 $\pm$ 18 $\mu$Jy beam$^{-1}$ and 814 $\pm$ 84 $\mu$Jy, respectively. 
This high S/N is well above the threshold of 10 required to make a robust size measurement \citep{2018ApJ...854...97D,2018ApJ...866..159V}. 
The tidal interaction between the host galaxies would have stretched the star-forming regions, 
and this intrinsic size is at the extreme of the FIR size distribution of $z \gtrsim 6$ quasars observed so far \citep{2019PASJ...71..111I,2020ApJ...904..130V}. 

From the measured continuum fluxes, we determined the FIR luminosity ($L_{\rm FIR}$; 42.5--122.5 $\micron$) 
and the total IR luminosity ($L_{\rm TIR}$; 8--1000 $\micron$) 
assuming an optically thin modified black-body spectrum and values of dust temperature ($T_{\rm dust}$) and emissivity index $\beta$. 
Note that previous studies on $z > 6$ luminous quasars with starburst (SFR $\gg 100~M_\odot$ yr$^{-1}$) nature 
have typically adopted canonical values of $T_{\rm dust} = 47$ K and $\beta = 1.6$, based on the averaged 
IR spectral energy distribution of high redshift optically-luminous quasars \citep{2006ApJ...642..694B,2013ApJ...772..103L}. 
Recent ALMA Band 9 observations toward one of the most luminous quasars at $z > 6$, J0100$+$2802 at $z = 6.3$, 
revealed its $T_{\rm dust}$ as 48 K that is similar to the above value, although its $\beta$ is as high as 2.6 \citep{2023ApJ...946L..45T}.

However, it seems inappropriate to use such high $T_{\rm dust}$ for the galaxies likely in an earlier evolutionary phase, 
where AGN and star formation are both much less active than the averaged sample of, for example, \citet{2006ApJ...642..694B}. 
Hence, for the time being until we are constrained by multi-band observations, 
we assume $T_{\rm dust} = 35$ K, a typical value in local LIRGs \citep[interacting galaxies are usually categorized as LIRGs at $z \sim 0$, e.g.,][]{2012ApJS..203....9U}. 
This value (35 K) is still within a range of $T_{\rm dust}$ observed in $z \sim 6$ main-sequence (normal) galaxies \citep{2021MNRAS.508L..58B}. 

We also correct for the contrast and the additional heating effects 
of the cosmic microwave background (CMB) radiation \citep{2013ApJ...766...13D} in Table \ref{tbl1}. 
Dust mass is estimated by adopting a rest-frame mass absorption coefficient of 
$\kappa_\lambda = 0.77 ({\rm 850 \micron}/\lambda)^\beta$ cm$^2$ g$^{-1}$ \citep{2000MNRAS.315..115D}. 
We emphasize that these results are sensitive to the assumed values of $T_{\rm dust}$ and $\beta$, 
which are known to be different in different sources \citep{2013ApJ...772..103L,2018ApJ...866..159V,2019MNRAS.489.1397L}
\footnote{For example, if we adopt $T_{\rm dust} = 47$ K, the resultant $L_{\rm TIR}$ and SFR will become $\sim 2.5\times$ higher than the values in Table \ref{tbl1}.}. 
The resultant value of $L_{\rm [CII]}/L_{\rm FIR}$, which is sensitive to the physical state of the ISM \citep[e.g.,][]{2013ApJ...774...68D,2018ApJ...861...95H}, is $7.7 \times 10^{-3}$. 
This is a factor $\sim 2$ higher than the value for the Milky Way \citep{2013ARA&A..51..105C}, 
but still within the range of nearby LIRGs \citep{2013ApJ...774...68D}. 

By assuming that the TIR continuum emission originates solely from star-forming regions, 
we estimate the SFR$_{\rm TIR}$ in Table \ref{tbl1} with the conversion ${\rm SFR_{TIR}} = 1.49 \times 10^{-10} L_{\rm TIR}/L_\odot$ \citep{2011ApJ...737...67M}. 
This conversion is based on the Kroupa initial mass function \citep[IMF,][]{2001MNRAS.322..231K}, but is adjustable for different IMFs if necessary. 
Although some works on luminous quasars claim the existence of quasar-heated dust at rest-FIR wavelengths \citep[e.g.,][]{2014ApJ...785..154L,2021ApJ...914...36I,2023MNRAS.523.4654T}, 
that effect may not be significant for these low-luminosity quasars. 
We also found that there is a significant discrepancy between SFR$_{\rm [CII]}$ and SFR$_{\rm TIR}$ particularly at C2 + Bridge and the Tail. 
Although one possible explanation is that the actual $T_{\rm dust}$ is much higher than the adopted 35 K, 
it is also plausible that dust-unobscured star formation dominates the total SFR at least at C2 + Bridge and the Tail, and would contribute significantly at C1, 
as [\ion{C}{2}] is sensitive to both obscured and unobscured star formation. 
Future multi-wavelength observations of the host galaxy-scale star formation, such as provided by JWST, will fully characterize the star forming activity of these host galaxies.

\subsection{Global gas dynamics}\label{sec3.3}
\begin{figure*}
\begin{center}
\includegraphics[width=\linewidth]{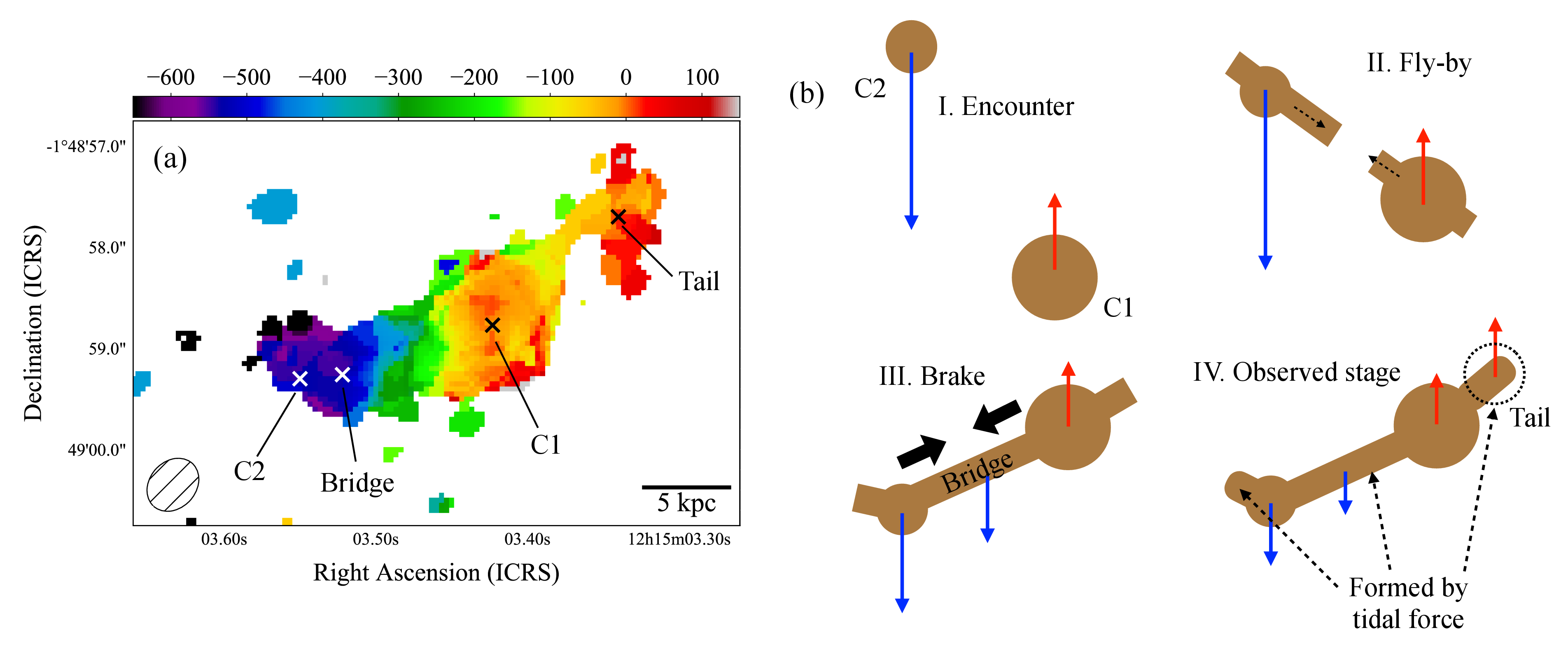}
\caption{
(a) Intensity-weighted mean [\ion{C}{2}] velocity field map of the pair system. 
The positions of C1, C2, Bridge, and Tail are marked by the crosses (see also Figure \ref{fig1}b). 
The bottom-left ellipse indicates the synthesized beam. 
The [\ion{C}{2}] line center measured at C1, with $1''.5$ circular aperture (Table \ref{tbl1}), is used as the systemic redshift. 
(b) Schematic explanation of the galaxy-galaxy interaction of the pair system. 
Four stages (encounter, fly-by, brake, and the observed stage) are illustrated. 
}
\label{fig3}
\end{center}
\end{figure*}

Figure \ref{fig3} shows an intensity-weighted mean [\ion{C}{2}] velocity map (moment-1) 
generated with the CASA task \verb|immoments| with $3.5\sigma$ clipping to avoid noisy pixels. 
As we already saw in the channel maps (Figure \ref{fig2}), there is a clear, {\it global} velocity gradient over the system, on scales of $\sim 4''$. 
Similar velocity gradients have also been observed in some quasars on a few kpc scales \citep[e.g.,][]{2016ApJ...816...37V,2018ApJ...854...97D,2021ApJ...914...36I}, 
as well as in PJ308$-$21 \citep[a pair of merging galaxies that hosts a quasar,][]{2019ApJ...880..157D}. 

This global dynamics can be explained by the following steps of tidal disruption of a satellite galaxy 
(host galaxy of C2) in close interaction with the heavier host galaxy of C1 (Figure \ref{fig3}b). 
Note that our attempt here is to explain the global behavior in a qualitative manner, rather than to fit the data with a physical model. 
We also note that the actual gravitational center of the host galaxy of C2 remains unclear (\S~\ref{sec3.1}). 
Future JWST observations will elucidate the mass distribution of the system and the dynamical nature more properly. 
\begin{itemize}
\item[I.] Close encounter: Two galaxies hosting C1 and C2 start to encounter. 
\item[II.] Fly-by: The galaxies are passing by one another. Gaseous material starts to be expelled from the two galaxies due to tidal interaction, 
which will eventually form the Bridge structure between them, as well as extended structures at the opposite sides of the galaxies (i.e., Tail). 
\item[III.] Braking phase: This process happens due to the gravitational deceleration and energy dissipation of the two galaxies. 
\item[IV.] The observed stage: Both of the Bride and the Tail are well developed due to the tidal force between the galaxies. 
\end{itemize}
This situation is very similar to what has been observed in PJ308$-$21 \citep{2019ApJ...880..157D}, 
and this close interaction indicates that the two galaxies will merge into one in a short time. 

The actual masses of the host galaxies of C1 and C2, and their ratio, 
are hard to estimate at this moment given the highly dynamic nature of the system. 
Nevertheless, we crudely estimate the dynamical mass ($M_{\rm dyn}$) of these host galaxies, 
by following the standard description in studies of $z > 6$ quasars \citep[e.g.,][]{2016ApJ...816...37V,2015ApJ...801..123W,2018PASJ...70...36I}. 
Here, the [\ion{C}{2}] emission of these galaxies is assumed to be mainly originating from thin rotating disks. 
The inclination angles ($i$) of the disks are determined from the axis ratios of the beam-deconvolved Gaussian fits to the emitting regions. 
We used the CASA task \verb|imfit| for this purpose, and found the beam deconvolved major ($a_{\rm maj}$) and minor ($a_{\rm min}$) axis sizes as 
$a_{\rm maj} \times a_{\rm min} = (0''.96 \pm 0''.09) \times (0''.68 \pm 0''.07)$ and $i = 45\arcdeg$ for C1, 
and $(0''.85 \pm 0''.26) \times (0''.35 \pm 0''.17)$ and $i = 66\arcdeg$ for C2 + Bridge regions, respectively. 
Although the latter was measured within a $1''$ box placed at the midpoint of C2 and the Bridge, we decided to use this to represent the host galaxy of C2, 
considering the spatial distribution of the [\ion{C}{2}] emission around there. 
The disk size is approximated as $D = 1.5 \times a_{\rm maj}$ to account for spatially extended low-level emission, 
and the circular velocity as $V_{\rm circ} = 0.75 \times {\rm FWHM}_{\rm [CII]}/\sin i$, 
where FWHM$_{\rm [CII]}$ is the Gaussian FWHM of the [\ion{C}{2}] line profile measured on the region 
(594 km s$^{-1}$ for C1 and 361 km s$^{-1}$ for C2 + Bridge). 
With these, we estimate that $M_{\rm dyn} \sim 9 \times 10^{10}~M_\odot$ for C1, 
and $\sim 5 \times 10^{10}~M_\odot$ for C2, respectively. 
Given these estimates, we found that their $M_{\rm dyn}$ are comparable to $M_{\rm H2}$, 
suggesting that this is a major merger system of gas-rich galaxies. 
The comparable $M_{\rm BH}$ of C1 and C2 \citep[$\sim 10^8~M_\odot$,][]{Matsuoka2024} also supports this scenario.

\subsection{Potential indication of AGN feedback}\label{sec3.4}
\begin{figure*}
\begin{center}
\includegraphics[width=\linewidth]{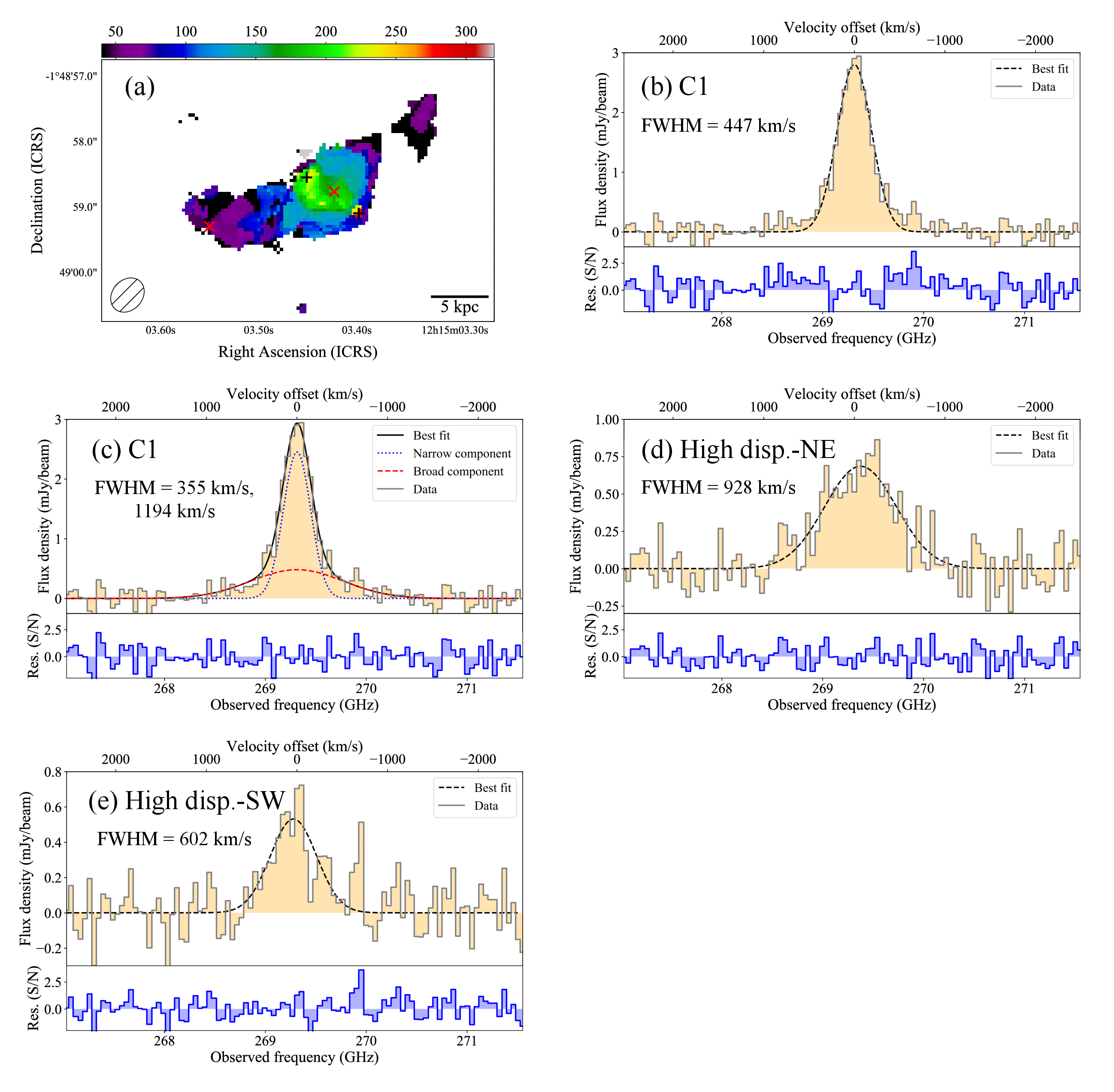}
\caption{
(a) Intensity-weighted [\ion{C}{2}] velocity dispersion map of the pair system. 
The two crosses indicate the locations of C1 and C2, whereas the two plus signs indicate the representative location of the high dispersion region (NE and SW). 
(b) The [\ion{C}{2}] spectrum measured at C1 with a single Gaussian fit. 
(c) Same as (b) but with a result of double Gaussian fit. 
(d) The [\ion{C}{2}] spectrum measured at the high dispersion region NE with a single Gaussian fit. 
(e) Same as (d) but for the case of the high dispersion region SW. 
In each panel of the line spectrum, residuals after subtracting the model from the data, normalized by the rms value (0.14 mJy beam$^{-1}$), 
together with resultant FWHM of the Gaussian fit, are also shown. 
}
\label{fig4}
\end{center}
\end{figure*}

Figure \ref{fig4}a shows the intensity-weighted [\ion{C}{2}] velocity dispersion map 
made with the same CASA immoments task with $3.5\sigma$ clipping: 
if the line profile is purely Gaussian, this dispersion indicates the Gaussian sigma.  
Overall, the dispersion ranges from 50 to 100 km s$^{-1}$ over the system, 
which is comparable to nearby merging galaxies \citep[e.g.,][]{2015ApJ...803...60S,2018ApJ...863..143C}. 
On the other hand, we found globally elevated dispersions around C1 ($\sim 100-150$ km s$^{-1}$), 
as well as linearly-distributed high dispersion regions from north-east to south-west across C1. 
We specified two representative positions along this line, marked with plus signs in Figure \ref{fig4}a, 
which are referred to as High Dispersion North-East (NE) and South-West (SW) in the following. 
As these regions are distributed roughly symmetrically to C1, and the regions of high dispersion are relatively localized, 
we speculate that this is a hint of two-directional (e.g., bi-conical) AGN-driven outflows, 
rather than violent shocks due to the merger of galaxies \citep{2016ApJ...816L...6D,2018Sci...362.1034D,2020ApJ...890..149T}. 

We first inspected the [\ion{C}{2}] line spectrum of C1 (Figure \ref{fig4}b), which was measured with a single synthesized beam. 
We first fit the spectrum with a single Gaussian model, which returned (peak amplitude, center, FWHM) 
= ($2.8 \pm 0.1$ mJy beam$^{-1}$, $269.313 \pm 0.005$ GHz, $447 \pm 12$ km s$^{-1}$). 
It is intriguing that there is excess emission from the single Gaussian up to $\pm 1000$ km s$^{-1}$, 
which is also seen in the high velocity components in the channel maps (Figure \ref{fig2}). 
Thus, we next fit the data with a double Gaussian model (Figure \ref{fig4}c) by fixing the centroid of the second Gaussian to that of the first Gaussian, 
and found that the narrow core component has (peak amplitude, center, FWHM) = ($2.5 \pm 0.3$ mJy beam$^{-1}$, $269.314 \pm 0.005$ GHz, $355 \pm 21$ km s$^{-1}$), 
and the broad wing component has (peak amplitude, center, FWHM) = ($0.48 \pm 0.11$ mJy beam$^{-1}$, 269.314 GHz (fixed), $1194 \pm 193$ km s$^{-1}$). 
The addition of this second broad Gaussian is assessed with the standard $\chi^2$ value: 
the single Gaussian fit returns $\chi^2$/d.o.f. = 85.5/56 = 1.53 whereas the double Gaussian fit returns $\chi^2$/d.o.f. = 42.7/54 = 0.79 
when evaluated over the velocity range of $\pm 1500$ km s$^{-1}$. 
Hence, the double Gaussian fit is statistically preferred. 
Furthermore, we found that the line profiles measured at the high dispersion regions are unusually broad as well 
(FWHM = $928 \pm 72$ km s$^{-1}$ at NE, $602 \pm 75$ km s$^{-1}$ at SW), 
although it is well-fit by a single Gaussian profile (Figure \ref{fig4}d,e).  

The nearly symmetric appearance of the broad wing at C1 disfavors nearby (unseen) companions as its physical origin: 
multiple companions at a range of velocity offsets would be required within this single synthesized beam. 
Hence we hereafter interpret this broad wing as a potential indication of fast [\ion{C}{2}] outflows. 
We emphasize that such fast neutral outflows (including [\ion{C}{2}] and OH 119 $\micron$) in quasars 
have been observed in only a few cases at $z > 6$ \citep[e.g.,][]{2021ApJ...914...36I,2024ApJ...962....1S}. 
Note that, however, further higher angular resolution, higher sensitivity observations are definitely required to confirm the presence of [\ion{C}{2}] outflows. 

Assuming that the wing has appeared due to outflows, we roughly estimate some basic physical properties. 
First, the line flux of the broad component at C1 ($0.61 \pm 0.17$ Jy beam$^{-1}$ km s$^{-1}$) corresponds to $L_{\rm [CII]} = (5.83 \pm 0.16) \times 10^8~L_\odot$. 
The outflowing atomic mass in neutral hydrogen gas \citep{2010ApJ...714L.162H,2020A&A...633A..90G} is 
\begin{equation}
\begin{split}
\frac{M_{\rm out}}{M_\odot} &= 0.77 \left( \frac{0.7 L_{\rm [CII],broad}}{L_\odot} \right) \left( \frac{1.4 \times 10^{-4}}{X_{\rm C^+}} \right) \\
& \times \frac{1 + 2e^{-91/T_{\rm ex}} + n_{\rm crit}/n}{2e^{-91/T_{\rm ex}}}, 
\end{split}
\end{equation}
where $X_{\rm C^+}$ is the ratio of C$^+$ abundance to H, $T_{\rm ex}$ is the gas excitation temperature in K, 
$n$ is the gas volume density in cm$^{-3}$, and $n_{\rm crit}$ is the critical density of the line ($\sim 3 \times 10^3$ cm$^{-3}$). 
We assumed that 70\% of the [\ion{C}{2}] emission originates from neutral photodissociation regions \citep{1997ARA&A..35..179H}. 
Following previous studies on [\ion{C}{2}] outflows \citep[e.g.,][]{2012MNRAS.425L..66M,2020A&A...633A..90G,2021ApJ...914...36I}, 
we estimated $M_{\rm out} = (5.6 \pm 1.5) \times 10^8~M_\odot$ for $X_{\rm C^+} = 1.4 \times 10^{-4}$ and $T_{\rm ex} = 200$ K, in the high density limit ($n \gg n_{\rm crit}$). 
As the outflows may smoothly propagate from the quasar outward along the linear regions of high dispersion (Figure \ref{fig4}a), 
here we measure the values within the single synthesized beam placed at C1. 
This gives the characteristic size (spatial extent) of $R_{\rm out} = 1.5$ kpc. 
Assuming an outflow velocity of $V_{\rm out} = {\rm FWHM}/2 = 597 \pm 97$ km s$^{-1}$ which stays constant throughout the flow \citep{2020A&A...633A.134L}, 
we obtain a neutral outflow rate of 
$\dot{M}_{\rm out, neutral} = M_{\rm out}V_{\rm out}/R_{\rm out} = 229 \pm 73~M_\odot$ yr$^{-1}$. 
This is comparable to the SFR$_{\rm [CII]}$ measured at C1 (Table \ref{tbl1}). 

\citet{2019MNRAS.483.4586F} studied multiphase outflows in nearby star-forming galaxies and AGNs, 
and found that the total outflow rate $\dot{M}_{\rm out,total}$, including the cold molecular phase, is typically $3 \times$ greater than the atomic-only value. 
Following this, we speculate that the actual mass loading factor $\eta = \dot{M}_{\rm out,total}/{\rm SFR}$ \citep{2020A&ARv..28....2V} would be $\gtrsim 2-3$ 
when we adopt the [\ion{C}{2}]-based SFR at C1. 
This indicates that a significant quenching of star formation is happening, while the gas depletion time is 
still as long as $\sim 3 \times 10^8$ yr, as compared to $\sim 10^7$ yr scales found in other quasars 
\citep[e.g.,][]{2024ApJ...962....1S}, owing to the rich amount of cold gas in the system. 

The driver of this feedback remains unclear as starburst-driven outflows typically show $\eta \sim 1-3$ \citep[e.g.,][]{2014A&A...562A..21C,2018MNRAS.473.1909G,2020A&A...633A..90G} 
and AGN-driven outflows have $\eta \gtrsim 5$ \citep[e.g.,][]{2014A&A...562A..21C,2017A&A...601A.143F,2019MNRAS.483.4586F}: our resultant $\eta$ is thus intermediate. 
We note, however, it is rare at any redshift for starburst-driven outflows to reach $V_{\rm out} > 500$ km s$^{-1}$ \citep[e.g.,][]{2005ApJ...621..227M,2020A&A...633A..90G}. 
Supposing the factor of 3 difference from the neutral outflow rate to the total outflow rate, we compute its kinetic energy as $\sim 2.6 \times 10^{43}$ erg s$^{-1}$. 
This is only 0.4\% of the quasar bolometric luminosity of C1 \citep{Matsuoka2024}, and is much smaller than 
the value expected in the blast energy-conserving AGN feedback frequently invoked in co-evolution scenario, 
even if it exists, to quench star formation of a host galaxy \citep[$\sim 5\%$,][]{2014MNRAS.444.2355C,2015ARA&A..53..115K}.

\section{Discussion and Summary}\label{sec4}
We have performed ALMA [\ion{C}{2}] line and FIR continuum emission observations 
toward the $z = 6.05$ quasar-quasar pair recently discovered by our Subaru/HSC observations \citep{Matsuoka2024}. 
These quasars are faint at rest-UV ($M_{\rm 1450} \gtrsim -23$ mag) with modest values of SMBH mass ($\sim 10^8~M_\odot$), 
but we detected very bright [\ion{C}{2}] emission, demonstrating that the system is rich in gas. 
Indeed, we estimated the total H$_2$ mass of the system as $\sim 10^{11}~M_\odot$ (or at least on the order of $10^{10}~M_\odot$) from the [\ion{C}{2}] luminosity. 
The masses of the host galaxies, crudely estimated from the [\ion{C}{2}] spatial distributions and line widths, are on the same order, showing that the system is very gas-rich. 
The [\ion{C}{2}]-based total SFR of the system is $\sim 550~M_\odot$ yr$^{-1}$ (dust-obscured SFR traced by the IR continuum emission is $\sim 100~M_\odot$ yr$^{-1}$). 
The large amount of gas sustains this high star-forming activity, which is most likely triggered by the galaxy-galaxy interaction. 

Recent simulations predict an increase in the fraction of SMBH-SMBH pairs with increasing redshift due to higher rates 
of galaxy-galaxy mergers \citep[e.g.,][]{2016MNRAS.460.2979V,2022MNRAS.514..640V,2019NewAR..8601525D}. 
ALMA observations of $z \gtrsim 5$ quasars have revealed the presence of companion galaxies (separation $\lesssim 50$ kpc) 
around up to 30--50\% of the systems \citep{2017Natur.545..457D,2017ApJ...836....8T,2017ApJ...850..108W,2019ApJ...882...10N,2020ApJ...895...74N}. 
Recent JWST observations have started to reveal close companion galaxies or mergers in many high-$z$ quasars, 
which were overlooked in previous works due to insufficient resolutions \citep[e.g.,][]{2023A&A...678A.191M,2023A&A...679A..89P}. 
These works demonstrate the importance of mergers of galaxies as the triggering mechanism of luminous quasars at high redshift. 
These are also in line with the recent hydrodynamic simulations that predict $z \sim 6$ quasars are a part of complex, 
gas- and dust-rich merging systems containing multiple sources \citep{2021MNRAS.503.2349D,2022MNRAS.514.1672V}. 

However, while some JWST observations have also uncovered dual AGNs (i.e., mergers) 
in lower luminosity ($L_{\rm Bol} \lesssim 10^{45}$ erg s$^{-1}$) and lower BH mass ($M_{\rm BH} \lesssim 10^7~M_\odot$) populations 
\citep[e.g.,][]{2023ApJ...959...39H,2023arXiv230801230M,2023arXiv231203589U}, 
previous works on massive BH pairs with $M_{\rm BH} \gtrsim 10^8~M_\odot$ were solely performed for luminous quasars with $M_{\rm 1450} \lesssim -26$ mag. 
Our observations now probe a much lower luminosity pair of quasars ($M_{\rm 1450} \gtrsim -23$ mag), 
which can serve as a benchmark of this unexplored regime of low-luminosity, but massive BH of $\sim 10^8~M_\odot$. 
It is interesting that AGN feedback might be happening already in this early stage (\S~\ref{sec3.4}), although its impact on the host galaxy would yet be small. 
Hence, powerful AGN feedback typically invoked in galaxy evolution models, if exists, will happen at later evolutionary phases 
(i.e., coalescent phase or UV-bright quasar phase).

According to models of merger-driven galaxy evolution \citep[e.g.,][]{1988ApJ...325...74S,2005Natur.433..604D,2006ApJS..163....1H,2019MNRAS.488.4004L}, 
both star formation and AGN are activated by the interaction of gas-rich galaxies. 
But the activity is further boosted by factors of 10--100 when the actual coalescence/merger happens \citep{2008ApJS..175..356H,2022ApJ...936..118Y}. 
Thus, we expect that this pair will evolve to a luminous quasar ($M_{\rm 1450} \lesssim -26$ mag) with high SFR of $> 1000~M_\odot$ yr$^{-1}$, 
comparable to the value for optically luminous (i.e., SDSS-class) quasars observed so far at high redshifts \citep[e.g.,][]{2018ApJ...866..159V,2021A&A...645A..33B}. 
Such evolution can be sustained by the rich amount of ISM available in this system (Table \ref{tbl1}). 
That is, this pair system may well correspond to a progenitor of the most luminous quasars in the early universe. 
Future multiwavelength observations, including JWST measurements of the stellar mass distribution and ionized outflows 
\citep[see examples in, e.g.,][]{2023A&A...678A.191M,2023Natur.621...51D}, 
aided with dedicated hydrodynamic simulations, will characterize the nature of this system, and allow us to predict its fate in greater detail.

\begin{acknowledgments}
We appreciate the anonymous reviewer's very constructive comments to improve this manuscript. 
This paper makes use of the following ALMA data: 
ADS/JAO.ALMA\#2019.A.00019.S. 
ALMA is a partnership of ESO (representing its member states), 
NSF (USA) and NINS (Japan), together with NRC (Canada), 
NSTC and ASIAA (Taiwan), and KASI (Republic of Korea), 
in cooperation with the Republic of Chile. 
The Joint ALMA Observatory is operated by ESO, AUI/NRAO, and NAOJ. 
We appreciate the ALMA Director for granting the observing time for the DDT program presented here. 

T.I., Y.M., Y.T., M.I., H.U., K.K. are supported by Japan Society for the Promotion of Science (JSPS) KAKENHI Grant Number 
JP23K03462, JP21H04494, JP23K22537, JP21K03632, [JP20H01953, JP22KK0231, JP23K20240], and [JP17H06130, JP22H04939, JP23K20035], respectively. 
K.I. acknowledges support under the grant PID2022-136827NB-C44 provided by MCIN/AEI/10.13039/501100011033 / FEDER, UE. 
T.I. is supported by the ALMA Japan Research Grant for the NAOJ ALMA Project, NAOJ-ALMA-319. 
\end{acknowledgments}

\appendix
\section{[\ion{C}{2}] line spectra}\label{App-I}
\begin{figure}
\begin{center}
\includegraphics[width=\linewidth]{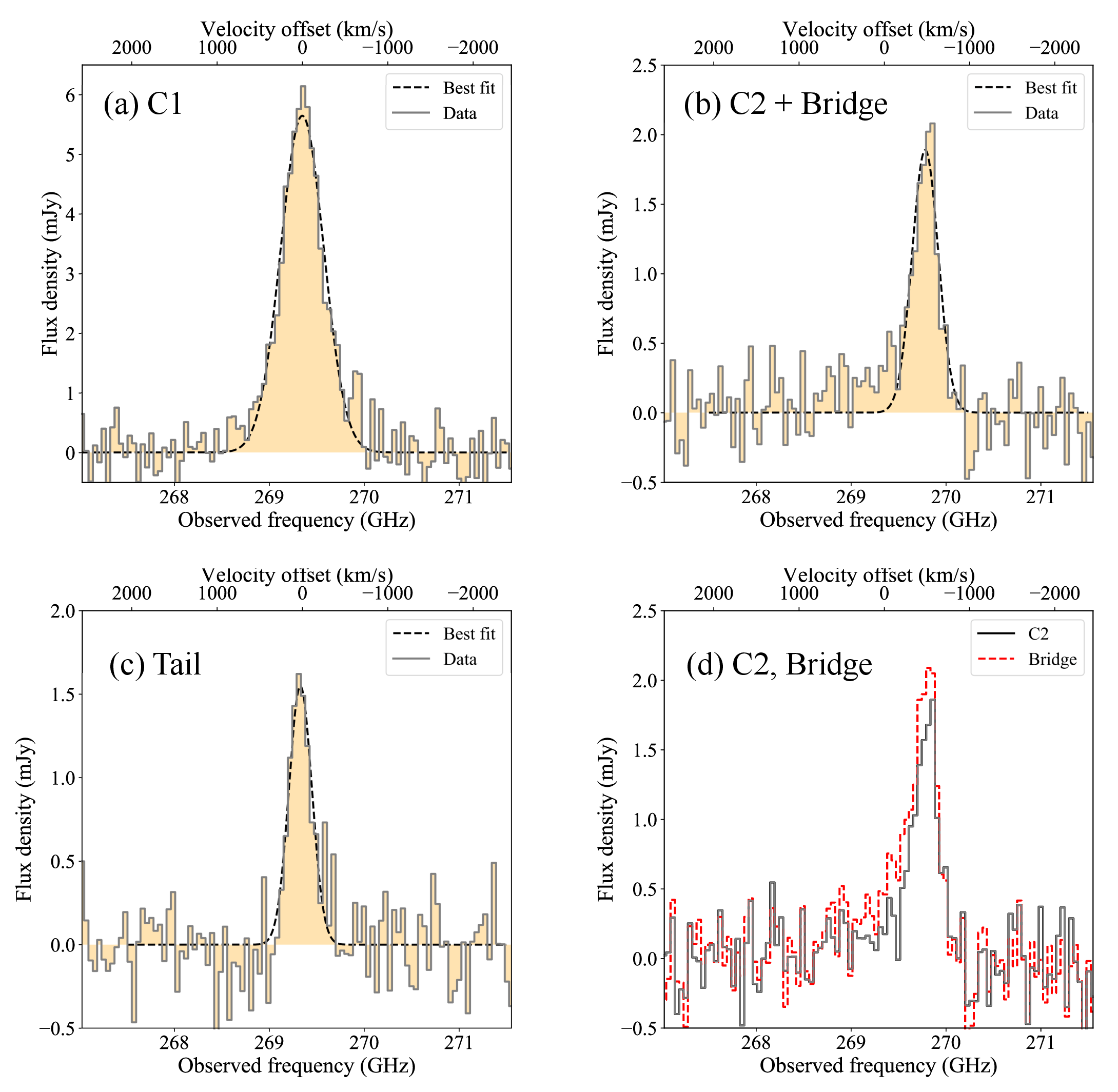}
\caption{
(a) [\ion{C}{2}] line spectrum measured at the position of C1, with $1''.5$ aperture. The $1\sigma$ rms noise level is 0.395 mJy. 
(b)(c) [\ion{C}{2}] line spectra of C2 + Bridge and Tail measured with a common $1''.0$ aperture. 
The $1\sigma$ noise level for these spectra is 0.254 mJy. 
In the panels (a)(b)(c), the results of single Gaussian fits are also shown (see Table \ref{tbl1} for the results). 
(d) Individual [\ion{C}{2}] spectra of C2 and the Bridge measured with a $1''.0$ aperture. 
As the spatial separation between these two positions are small, the measured values here are degenerated. 
}
\label{fig_App1}
\end{center}
\end{figure}

The [\ion{C}{2}] line spectra measured at the positions of C1, C2, Bridge, Tail (see Figure \ref{fig1}), as well as that of C2 + Bridge, are presented in Figure \ref{fig_App1}. 
Results of our single Gaussian fits to these spectra are summarized in Table \ref{tbl1}.

\section{Smoothed velocity channel maps}\label{App-II}

\begin{figure}
\begin{center}
\includegraphics[width=\linewidth]{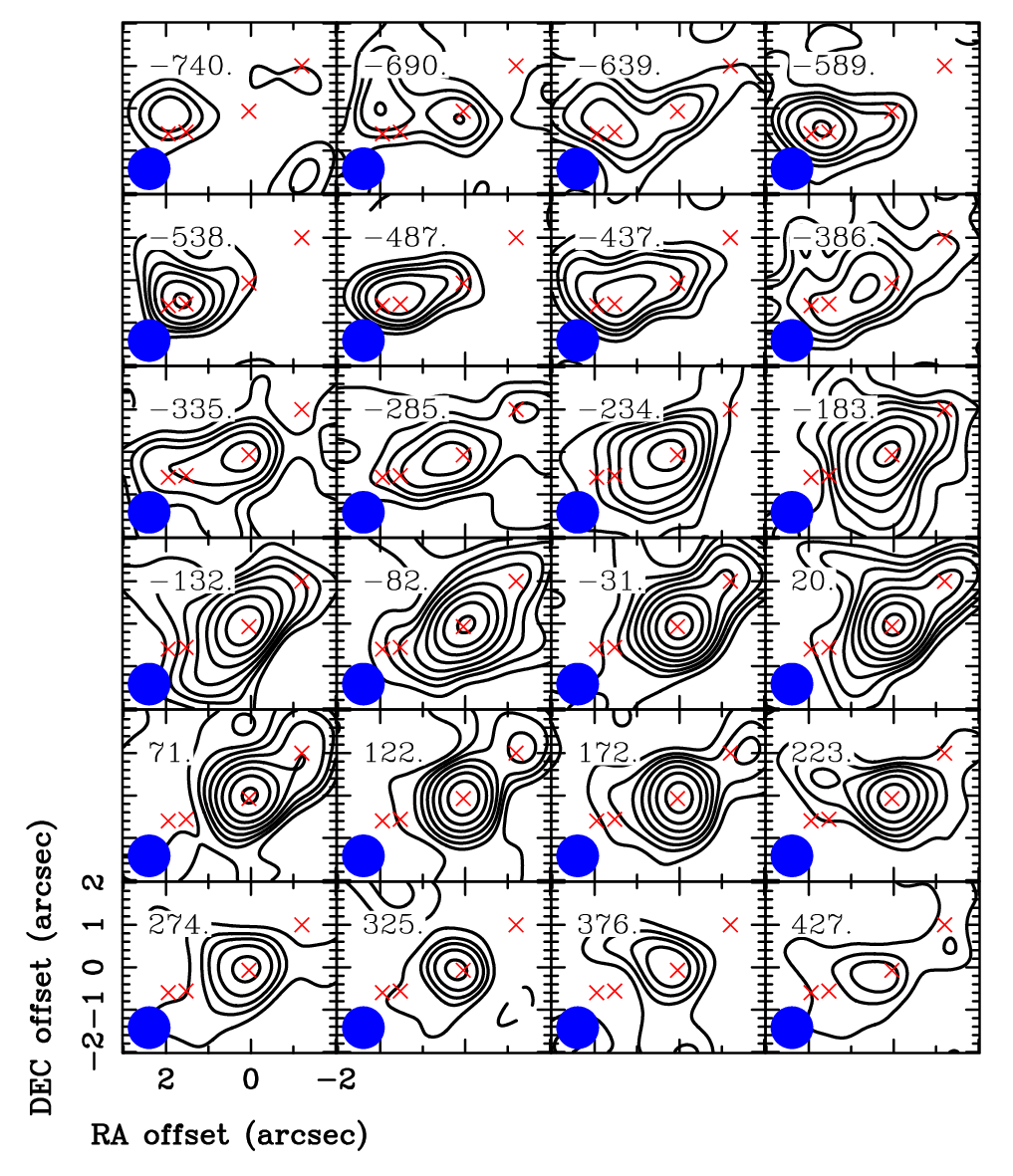}
\caption{
Velocity channel maps of [\ion{C}{2}] line of the pair system. The maps are made after smoothing the data to $1''.0$ resolution. 
The first contours indicate $\pm 1\sigma$ to enhance the faint emission, while the remaining are drawn from $\pm 2\sigma$ with an increment of $\sqrt{2}$, 
where $1\sigma = 0.254$ mJy beam$^{-1}$. The bottom-left ellipse in each panel indicates the synthesized beam ($1''.0$). 
The four positions (C1, C2, Bridge, Tail) are also marked (see Figure \ref{fig1}b). 
}
\label{fig_App2}
\end{center}
\end{figure}

\bibliography{Izumi2024a_ref}{}
\bibliographystyle{aasjournal}

\end{document}